
\magnification=1200
\baselineskip=12pt
\settabs 20 \columns
\+&&&&&&&&&&&&&&& UM-TH-94-33 \cr
\+&&&&&&&&&&&&&&& hep-ph/9411292 \cr
\+&&&&&&&&&&&&&&& October  , 1994 \cr

\vskip2truecm

\centerline{\bf ${1\over m_b}$ and ${1\over m_t}$ Expansion of the
 Weak Mixing Matrix}

\vskip1.5truecm

\centerline {York-Peng Yao}
\medskip

\centerline { \sl Randall Laboratory of Physics, University of
Michigan }
\smallskip

\centerline { \sl Ann Arbor, MI 48109 U. S. A. }

\vskip2truecm

\centerline{\bf Abstract}
\bigskip
We perform a $1/m_b$ and $1/m_t$ expansion of the Cabibbo-Kobayashi-
Maskawa mixing matrix.  Data suggest that the dominant parts of the
Yukawa couplings are factorizable into sets of numbers $\vert r>$,
$\vert s>$, and $\vert s'>$, associated, respectively, with the
left-handed doublets, the right-handed up singlets, and the right-
handed down singlets.  The first order expansion is consistent with
Wolfenstein parameterization, which is an expansion in $sin \theta _c$
to third order.  The mixing matrix elements in the present approach
are partitioned into factors determined by the relative orientations
of $\vert r>$, $\vert s>$, and $\vert s'>$ and the dynamics provided
by the subdominant mass matrices.  A short discussion is given of
some experimental support and a generalized Fritzsch model is used
to contrast our approach.

\bigskip
\noindent
PACS number(s): 12.15.Ff, 12.50.Ch
\vfill \eject
\bigskip
It is not an overstatement that a very perplexing and yet most challenging
problem in particle physics lies in explaining the disparateness of
fermion masses and in some of
the off-diagonal elements of the Cabibbo-Kobayashi-Maskawa (CKM) mixing
matrix.  There has been a great amount of  activities in this area.
Generally speaking,
it has become somewhat of an art form to postulate certain textures [1]
in the quark mass matrices to explain these peculiarities.  While this
may well be a first step towards formulating some dynamical principle
and to fathom
a Nature's deep secret, we would like to take a different approach here.
Our emphasis is on the empirical fact of the strong decoupling of the
third family from the first two in the CKM matrix, which will not
be perceived as numerical accidents.
We shall provide a short discussion of the relevant experimental
evidence and analyse a generalized Fritzsch model for contrast
in Appendices.

\smallskip
Our starting point is an attempt to separate out the large from the
small at the very beginning.  By this, we mean to work in a framework
in which the mass of the top quark $m_t$ is {\it ab initio} taken as
the largest
scale in a theory with three families.  The mass of the bottom
quark $m_b$ is also initially built into our analysis as the heaviest
member of the down-type quarks [2].  For the CKM matrix, we shall regard
the mixing of the first two families predominantly due to approximate
mass degeneracy relative to $m_t$ and $m_b$  The magnitude of these
elements is taken to be of
order unity $O((1/m_t)^0,(1/m_b)^0)$.  The lifting of degeneracy
in $m_u\neq m_c$
and $m_d\neq m_s$ provides the dynamics for their eventual assumed values.
In the following,
we shall organize in a way which we perceive as natural for an expansion
in the small parameters.  We shall see that upon a conjecture to be
made shortly, an expansion to first order in $m_b^{-1}$ and $m_t^{-1}$
will give the same accuracy for the CKM matrix as in Wolfenstein
parameterization [3].

\smallskip
We can diagonalize, for example, the up-type quark mass matrix $M$ by
a biunitary transformation [4].  Let $\vert \hat r >$ and
 $<\hat s \vert $ be, respectively, the normalized right and left
eigenvector for the top quark.  Then we write
$$M=\bar m_t\vert \hat r > < \hat s \vert +\epsilon {\cal M},\eqno (1)$$
where $\epsilon {\cal M}$ can be written as a sum of two terms
pertaining to the up and the charm quark,
which have the corresponding structure as the first term.  However,
we shall refrain from doing that, because unless we know what the mass
matrix is, we would not know what these other eigenvectors are like in
relation to $\vert \hat r>$.  Fortunately, for our immediate purpose,
we need to note only that
$\epsilon {\cal M}$ is relatively small.  Similarly, for the down- type
quark mass matrix $M'$, we pick out the right $\vert \hat s'>$
and left $<\hat r'\vert$ eigenvectors for the bottom quark and write
$$M'=m_b\vert \hat r' > < \hat s' \vert +\epsilon '
\widetilde {\cal M'}.$$
We now make the following conjecture: $\vert \hat r>$ and $\vert \hat r'>$
are almost aligned, i.e. if we replace $\vert \hat r'>$ by $\vert \hat r>$
in $M'$, the difference, which will be put into $\epsilon ' \widetilde
{\cal M'}$, is small compared to $m_b$.  Thus,
$$M'=\bar m_b\vert \hat r > < \hat s' \vert +\epsilon '{\cal M'}.\eqno (2)$$
In the above, $\epsilon $ and $\epsilon '$ are counting parameters, which
will be set to unity after the count.
As we shall see, this momentarily gives small off-diagonal matrix elements
$V_{ts}, \ V_{td}, \ V_{ub},$ and $V_{cb}$.

\smallskip
One may forgo the presentation we just made, which is motivated from
simple mathematical consideration.  One takes note that the left-handed
up and down type quarks belong to the same doublets.
Our observation is equivalent to the proposal that the dominant piece
of the Yukawa coupling matrices are factorizable into
a factor $\sim \vert r>$ which is connected with the left-handed doublets
$Q_L=(U_L,D_L)$, a factor $\sim \vert s>$ with the right-handed up singlets
$U_R$, and another factor $\sim \vert s'>$ with the right-handed down
singlets $D_R$.  This
may hint at some exchange-type of mass generating mechanism (Fig. 1).
To accomodate this probable interpretation, and also to facilitate
writing expressions for the up-type and the down-type symmetrically,
we shall assume that $\vert \hat r>, \ \vert \hat s>$ and $\vert
\hat s'>$ in Eqs.(1) and (2) are not the exact eigenvectors for the top
and the bottom
fields.  There are $O(\epsilon , \ \epsilon ')$ corrections.  By
the same token, $\bar m_t$ differs from the true mass $m_t$ of the top
quark by $O(\epsilon )$, and $\bar m_b$ from $m_b $ by $O(\epsilon ').$

\smallskip
It proves convenient at this point to introduce a set of basis
vectors for the up and the down flavor spaces.  Because the CKM matrix
connects left-handed quarks, we need to deal with only $MM^\dagger$
and $M'M'^\dagger$.  For the up-sector, the basis vectors we choose
are $\vert \hat r>$,
$$\ \vert \hat t>\equiv(\vert \hat s >-<\hat r\vert
\hat s >\vert \hat r>)/(1-\vert <\hat r \vert \hat s >\vert ^2)^{1/2},$$
and
$$\vert \hat n>\equiv \vert \hat r^\star \times \hat s^\star >/
(1-\vert <\hat r \vert \hat s >\vert ^2)^{1/2}.\eqno (4)$$
Similar expressions
with $s \rightarrow s'$ will be used for $\vert \hat t'> $ and
$\vert \hat n'>$ for the down-sector.  It may seem odd that the intrinsically
right-handed vectors $\vert \hat s >$ and $\vert \hat s' >$ are also
used in the left spaces, but they are certainly the preferred vectors
and may in fact tell us something about the much discussed left-right
symmetry, or the lack thereof.

\smallskip
We take $MM^\dagger $ and $M'M'^\dagger$
and determine the masses and eigenvectors to an accuracy such that the
CKM matrix will be obtained to $O(\epsilon )$ or $O(\epsilon ')$.
For the up-sector, it is easy to arrive at the normalized mass eigenstates
$$\vert y_{u,c}>=\vert y^0_{u,c}>-(\epsilon /m_t)<\hat s\vert
{\cal M^\dagger } \vert y^0_{u,c}>\vert \hat r>,$$
$$\vert y_t >=\vert \hat r >+(\epsilon /m_t)<y^0_u\vert {\cal M}\vert \hat s>
\vert y^0_u>+(\epsilon /m_t)<y^0_c\vert {\cal M}\vert \hat s>\vert y^0_c>,
\eqno (5)$$
where $m_t$ can be equated with $\bar m_t$ to this order, and
$\vert y^0_{u,c}>$ are the two 0-th order eigenvectors for the up
and the charm quarks
$$\vert y^0_i>=\vert \hat t><\hat t\vert y^0_i>
+\vert \hat n><\hat n\vert y^0_i>,\eqno (6)$$
determined by the equations
$$<\hat t\vert {\cal H}\vert \hat t><\hat t\vert y^0_i>+
<\hat t\vert {\cal H}\vert \hat n><\hat n\vert y^0_i>
=\lambda _i<\hat t \vert y^0_i>,$$
and
$$<\hat n\vert {\cal H}\vert \hat t><\hat t\vert y^0_i>+
<\hat n\vert {\cal H}\vert \hat n><\hat n\vert y^0_i>
=\lambda _i<\hat n \vert y^0_i>. \eqno (7)$$
Here the dynamics is given by the subtracted 'Hamiltonian'
$$\eqalign {{\cal H}&={\cal M M^\dagger }-{\cal M}\vert \hat s><\hat s\vert
{\cal M^\dagger} \cr
&={\cal M }\vert \hat v ><\hat v \vert {\cal M^\dagger}
+{\cal M }\vert \hat n ><\hat n \vert {\cal M^\dagger}, \cr}$$
with
$$\vert \hat v>=<\hat s \vert \hat r>\vert \hat t>
-(1-\vert <\hat r\vert \hat s >\vert ^2)^{1/2}\vert \hat r>. \eqno (8)$$
Please note that the subspace which has been subtracted out is $\vert \hat s>$.
There can be leakage of dynamics from the $\underline {approximate}$ top state
$\vert \hat r>$ into the two low-lying
members.  We have taken the liberty to factor out a common factor
$\epsilon ^2$ in Eq.(6), i. e. $m_i^2=\epsilon ^2 \lambda _i, \ i= u, c.$
We can write down similar expressions for the down-type quarks with
the replacements $\epsilon \rightarrow \epsilon ', \ {\cal M
\rightarrow M'},$ etc.

\smallskip
For the CKM matrix, we just form the scalar products $V_{ij}=<y_{u,c,t}\vert
y'_{d,s,b}>.$  We shall adjust the phases so that
$$V_{ud}=<y^0_u\vert y'^0_d>+O(\epsilon ^2, \epsilon '^2, \epsilon \epsilon ')
= cos \theta _c,$$
$$V_{us}=<y^0_u\vert y'^0_s>+O(\epsilon ^2, \epsilon '^2, \epsilon \epsilon ')
= sin \theta _c,$$
$$V_{cd}=<y^0_c\vert y'^0_d>+O(\epsilon ^2, \epsilon '^2, \epsilon \epsilon ')
= -sin \theta _c,$$
and
$$V_{cs}=<y^0_c\vert y'^0_s>+O(\epsilon ^2, \epsilon '^2, \epsilon \epsilon ')
= cos \theta _c,\eqno (9)$$
in which $\theta _c$ stands for the Cabibbo angle.  These four elements
can be calculated from Eqs.(7)-(8) and the corresponding set for the d-s
system, once we postulate some dynamics for ${\cal H }$ and ${\cal H'}$.
The other elements are
$$V_{ub}=(\epsilon '/m_b)<y'_d\vert {\cal M'}\vert \hat s'>cos \theta _c
         +(\epsilon '/m_b)<y'_s\vert {\cal M'}\vert \hat s'>sin \theta _c
         -(\epsilon /m_t)<y_u\vert {\cal M}\vert \hat s>,$$
$$V_{td}=(\epsilon /m_t)<\hat s\vert {\cal M^\dagger}\vert y_u>cos \theta _c
         -(\epsilon /m_t)<\hat s\vert {\cal M^\dagger}\vert y_c>sin \theta _c
         -(\epsilon '/m_b)<\hat s'\vert {\cal M'^\dagger}\vert y'_d>,$$
$$V_{cb}=-(\epsilon '/m_b)<y'_d\vert {\cal M'}\vert \hat s'>sin \theta _c
         +(\epsilon '/m_b)<y'_s\vert {\cal M'}\vert \hat s'>cos \theta _c
         -(\epsilon /m_t)<y_c\vert {\cal M}\vert \hat s>,$$
$$V_{ts}=(\epsilon /m_t)<\hat s\vert {\cal M^\dagger}\vert y_u>sin \theta _c
         +(\epsilon /m_t)<\hat s\vert {\cal M^\dagger}\vert y_c>cos \theta _c
         -(\epsilon '/m_b)<\hat s'\vert {\cal M'^\dagger}\vert y'_s>,$$
and
$$V_{tb}=1+O(\epsilon ^2, \epsilon '^2, \epsilon \epsilon '), $$
from which it follows
$$V_{ub}\cong -V^\star_{td}cos \theta _c-V^\star_{ts}sin \theta _c,$$
and
$$V_{cb}\cong V^\star_{td}sin \theta _c-V^\star_{ts}cos \theta _c.
\eqno (10)$$
\smallskip
Let us be reminded that the expansion parameters in the present analysis
are $1/m_t$ and $1/m_b$.  One can easily check that unitarity holds to the
first order in the CKM matrix we just constructed.  Interestingly
enough, the Wolfenstein representation [3], which is obtained under the
assumption that $sin \theta _c$ is the only parameter of expansion
and which is accurate to $O((sin\theta _c)^3)$,
yields relations in Eq.(10) trivially.
Besides, the neglected terms in our approach are of order
$(m_s/m_b)^2$, which is $\approx O((sin \theta _c)^4)$.  This may not
be a numerical accident.
\smallskip
By design, all the matrix elements of ${\cal M}$ and ${\cal M'}$ should
be at most of order $m_c$ or $m_s$, respectively.  Thus, making use of the
fact that $m_s/m_b \gg m_c/m_t$, we have
$$V_{td,ts}\cong -(\epsilon '/m_b)<\hat s'\vert {\cal M'^\dagger}
\vert y'_{d,s}>,\eqno (11)$$
which provide the absolute normalization for the mixing of the
third to the first and the second families, $\sim m_d/m_b$
and $\sim m_s/m_b$, respectively.
\smallskip
We now discuss the mixing of the first two families, where
$$\eqalign {<y_{i=u,c}^0\vert y_{j=d,s}'^0>=&
<y_i^0\vert \hat t><\hat t'\vert y_j'^0><\hat t \vert \hat t'>
+<y_i^0\vert \hat n><\hat n'\vert y_j'^0><\hat t \vert \hat t'>^\star \cr
& +<y_i^0\vert \hat t><\hat n'\vert y_j'^0><\hat t \vert \hat n'>
-<y_i^0\vert \hat n><\hat t'\vert y_j'^0><\hat t \vert \hat n'>^\star. \cr}
\eqno (12)$$
This is composed of intra-space dynamics, determined by Eqs.(7)-(8) and
a similar set for the down-space, and interspace 'geometry'

$$<\hat t\vert \hat t'>=<\hat n'\vert \hat n >; \;
<\hat t\vert \hat n'>=-<\hat t'\vert \hat n >. \eqno (13)$$
At one extreme,
one model may be that $\vert \hat s>$ and $\vert \hat s'>$ are
aligned much better than $O((m_d/m_s)^{1/2}$.  Then, $<\hat t \vert
\hat t'>=1$ and $<\hat t \vert \hat n'>=0.$   A Fritzsch-type [5] model
$${\cal M}_{a=\hat t, \hat n \ b=\hat v, \hat n}=
\pmatrix {-(m_c \ - m_u) & (m_c m_u)^{1/2} & \cr
(m_c m_u)^{1/2} & 0 & \cr},$$

$${\cal M'}_{a=\hat t', \hat n' \ b=\hat v', \hat n'}=
\pmatrix {-(m_s \ - m_d) & (m_s m_d)^{1/2} & \cr
(m_s m_d)^{1/2} & 0 & \cr} \eqno (14)$$
will dictate the intraspace dynamics and lead to
$sin \theta _c\approx -(m_d/m_s)^{1/2}$.
\smallskip
At the
other extreme, we may have a model in which the 'Hamiltonian'
of Eq.(8) is already in the diagonal form, with e. g.
$$<y_u^0\vert \hat t>=1, \ <y_u^0\vert \hat n>=0; \
<y_c^0\vert \hat t>=0, \ <y_c^0\vert \hat n>=1; \ $$
$$<\hat t'\vert y_d'^0>=1, \ <\hat n'\vert y_d'^0>=0; \
<\hat t'\vert y_s'^0>=0, \ <\hat n'\vert y_s'^0>=1. \eqno (15)$$
This will give
$$cos \theta _c=<\hat t \vert \hat t'>, \ \ \ sin \theta _c
=<\hat t \vert \hat n'>, \eqno (16)$$
which are completely due to the geometric orientations of $\vert \hat r>,
\; \vert \hat s >$ and $\vert \hat s'>$.  Clearly, some clarifying
dynamical principle is awaited to give some credence.  Work is in
progress in this as well as to calculate $< \hat s'\vert
{\cal M'^\dagger }\vert y'_{d,s}>$ of Eq.(11).

\smallskip
In conclusion, we find data in suggestion that the left eigenvector of
the top quark is very much aligned with the left eigenvector of the
bottom quark.  This may be elevated to a conjecture that the
dominant Yukawa couplings factorize into a set of
numbers $\vert r >$  assigned to the left-handed quark
doublets $Q_L=(U_L,\ D_L)$, a set $\vert s >$ to the right-handed up
singlets $U_R$, and another set $\vert s' >$ to the right-handed down singlets
$D_R$.  The first order expansion in $1/m_t$ and $1/m_b$ is in
agreement with Wolfenstein parameterization, as indicated by Eq.(10).
This approach also sharpens the separation between the large
and the small, particularly in their dynamics and geometry.
\smallskip
{\it Added Note:} Professor A. Kagan has kindly pointed out to me
some earlier related work.$^{5'}$
\smallskip
This work has been supported partially by the U. S. Department of
Energy.
\vfill
\eject
\noindent
{\bf Appendices:}
\bigskip
\noindent
{\it Experimental data}
\bigskip
The experimental values of the CKM matrix elements have been summarized
by the Particle Data Group.$^6$  Let us look at them, particularly at those
obtained under the assumption of three families and constrained by
unitarity:
$$ (|V_{ij}|)	=
\pmatrix {0.9747 \ to \ 0.9759  & 0.218 \ to \ 0.224 &  0.002 \ to \ 0.005
& \cr
0.218 \ to \ 0.224 & 0.9738 \ to \ 0.9752 & 0.032 \ to \ 0.048 & \cr
0.004 \ to \ 0.015 & 0.030 \ to \ 0.048 & 0.9988 \ to \ 0.9995 &\cr} $$
{}From this table, we have:
\smallskip
\noindent
(i)$$\eqalign{ 1-\{ |V_{ud}||V_{cs}|&+|V_{us}||V_{cd}|\}
=1-\{0.9747\times 0.9738 +(0.224)^2\} \ \ \ or \cr
& =1-\{0.9757\times 0.9752 +(0.218)^2\} \cr
& \approx 0.0008, \cr}$$
\smallskip
\noindent
(ii) $$1-|V_{tb}|\approx 0.0012,$$
\smallskip
\noindent
(iii) $$|V_{ud}|^2-|V_{cs}|^2\ll (0.9759)^2-(0.9738)^2=0.0041.$$
These are various measures of how the third family decouples from
the first two.
While somewhat indirect, we interprete these to indicate that the
differences are all of order $({m_s\over m_b})^2$.  The situation
will be much improved when we have a better determination of
$|V_{cd}|=0.204\pm 0.017$ and $|V_{cs}|=1.01\pm 0.18$.
\smallskip
One can also relate these to the Wolfenstein paramterization
$$V=\pmatrix {1-\lambda ^2/2 & \lambda & \lambda ^3A(\rho - i\eta)& \cr
-\lambda & 1-\lambda ^2/2 & \lambda ^2 A & \cr
\lambda ^3A(1-\rho -i\eta) & -\lambda ^2A & 1 & \cr},$$
which is an expansion in the (assumed single) parameter
$\lambda =sin \theta _c$ to the
third order.  The neglected terms are of order $\lambda ^4\approx
({m_s\over m_b})^2$.
Taken together, the data correspond to:
\smallskip
\noindent
(i) $$V_{tb}=1+O(({m_s\over m_b})^2),$$
\smallskip
\noindent
(ii) $$V_{ud}=V_{cs}+O(({m_s\over m_b})^2),$$
\smallskip
\noindent
(iii) $$|V_{us}|=|V_{cd}|+O(({m_s\over m_b})^2),$$
(iv) $$|V_{td}|\approx |V_{ub}|\sim {m_d\over m_b}, \ \ \
|V_{ts}|\approx |V_{cb}|\sim {m_s\over m_b}.$$
\smallskip
We have used:$^7$
$${m_s\over m_b}(1Gev)={0.199 \over 7.005}=0.0284
={m_s\over m_b}(m_w)={0.087 \over 3.063},$$
$${m_d\over m_b}(1Gev)={0.0099\over 7.005}=0.0014
={m_d \over m_b}(m_w)={0.00433 \over 3.063}.$$
\vglue1in
\noindent
{\it Fritzsch-type models}
\bigskip
Clearly, there are far too many parameters in the mass matrices.  In
attempts to decrease the number, Fritzsch and others proposed to put
some of the elements to zero, which could lead to relations between
masses and mixing matrix elements.  This has been coined as 'texture
studies' or more poetically 'stitching the Yukawa quilt'.$^8$
\smallskip
Let us take a typical example to point out what needs to be done
and perhaps the potential troubles in all such models.  Here$^9$
$$M=\pmatrix {0 & x & 0 & \cr
x^\star & \alpha & b & \cr
0 & b^\star & a & \cr }, \ \ \
M'=\pmatrix {0 & y & 0 & \cr
y & \beta & f & \cr
0 & f & d & \cr}$$
where by phase choice, all the entries other than $x=|x|e^{i\delta _x}$
and $b=|b|e^{i\delta _b}$ are real.  The original Fritzsch model
corresponds to setting $\alpha = \beta =0$, which turns out to be
inconsistent with $m_t \approx 174 \ Gev$.  In order to enforce
the sequencing of eigenvalues $m_u < m_c < m_t$ and $m_d < m_s < m_b$, one has
$m_u -m_c < \alpha < m_t - m_c$ and $m_d - m_s < \beta < m_b -m_s$.
Furthermore, to retain some vestige of a radiative mass hierarchy
interpretation, which was
the motivation for the model, one should confine the ranges to
$|\alpha |<m_c$ and $|\beta |< m_s$.
\smallskip
For the decoupling indicators, one can easily obtain
\smallskip
\noindent
(i)$$ 1-\{ |V_{ud}||V_{cs}|+|V_{us}||V_{cd}|\} \approx {1\over 2}\Delta ,$$
\smallskip
\noindent
(ii) $$1-|V_{tb}|\approx {1\over 2} \Delta ,$$
\smallskip
\noindent
(iii) $$|V_{ud}|^2-|V_{cs}|^2\approx \Delta ,$$
where
$$\Delta =\Delta _t+\Delta _b-2\Delta _t^{1/2}\Delta _b^{1/2}cos \delta _b,$$
$$\Delta _b={\beta +m_s-m_d \over m_b}, \ \ \
\Delta _t={\alpha +m_c-m_u \over m_t}.$$
\smallskip
\noindent
In order to agree with data, we need
$${1\over 2}\Delta  \approx 2({m_s\over m_b})^2,$$
which requires a tuning of
$$\alpha \cong -{1\over 2}m_c \ \ \ \ \ \beta \cong -m_s.$$
We do not know of any symmetry which demands this.  We may remark
that once we decide to tune $\Delta $ to decrease to
$\approx O(({m_s\over m_b})^2)$, then there are other extra terms
to the same order for the right hand sides of the decoupling
indicators above. However, the simple predictions
$$|V_{(ub, \ cb, \ ts, \ td)}|\cong N_{(u, \ c, \ s, \ d)}\Delta ^{1/2},$$
where
$$N_{(u, \ c)}=\sqrt {m_{(u, \ c)} \over m_u + m_c}, \ \ \
N_{(d, \ s)}=\sqrt {m_{(d, \ s)} \over m_d + m_s}, $$
and
$$|V_{(ud, \ cs)}|\cong cos \theta _c , \ \ \
  |V_{(us, \ cd)}|\cong sin \theta _c ,$$
where
$$cos \theta _c =(N_u^2N_d^2+N_c^2N_s^2+2N_uN_cN_dN_scos\delta _x)^{1/2},$$
$$sin \theta _c =(N_u^2N_s^2+N_c^2N_d^2-2N_uN_cN_dN_scos\delta _x)^{1/2}.$$
still hold.  Corrections are
of order $(m_s/m_b)^2$ after the tuning.
\smallskip
{}From this brief exposition, it is evident that to save the radiative
mass hierarchy picture, it is necessary to tune some parameters,
$\alpha$ and $\beta $.  The sole purpose is to accelerate the
decoupling of the third family from the first two.  Our proposed
formulation incorporates this at the very start.
\vfill
\eject
\noindent
{\bf References:}
\smallskip
\noindent
\settabs 20 \columns
\+ [1] &See, e.g., A. Kusenko and R. Shrock, Phys. Rev. {\bf D49}, 4962 (1994)
for a brief review.\cr
\+ [2] & The emphasis is somewhat akin to that by P. H. Frampton and C.
Jarlskog, Phys. \cr
\+ & Letts. {\bf 154B}, 421 (1985).\cr
\+ [3] & L. Wolfenstein, Phys. Rev. Letts. {\bf 51}, 1945 (1983).\cr
\+ [4] & We do not use polar decomposition to make the mass matrices
hermitian as in Ref.(2). \cr
\+ & We do not want to phase away
the information contained in $\vert \hat s>$ and $\vert \hat s'>$. \cr
\+ [5] & H. Fritzsch, Phys. Letts. {\bf 70B}, 436 (1977). \cr
\+ [5'] & B. S. Balakrishna, Phys. Rev. Letts. {\bf 60}, 1602 (1988); \cr
\+ & B. S. Balakrishna, A. Kagan and R. N. Mohapatra, Phys. Letts.
{\bf 205B}, 345 (1988); \cr
\+ & A. Kagan, Phy. Rev. {\it D 40}, 173 (1989), Proc. of Johns Hopkins
Workshop on Current \cr
\+ & Problems in Particle Theory, 1991 (unpublished); \cr
\+ [6] & Particle Data Group, L. Montanet {\it et al.}, Phys. Rev. {\bf D50},
  1173 (l994). \cr
\+ [7] & Y. Koide, unpublished report US-94-05 (Oct. 1994).\cr
\+ [8] & P. Ramond, R. G. Roberts and G. G. Ross, Nucl. Phys. {\bf B406},
    19 (1993).\cr
\+ [9] & H. Fritzsch, Phys. Letts. {\bf B73}, 317 (1977),
Nucl. Phys. {\bf B155},189 (1979);\cr
\+ & L. F. Li, Phys. Letts. {\bf B84},461 (1979); \cr
\+ & S. N. Gupta and J. M. Johnson, Phys. Rev. {\bf D44}, 2110 (1991);\cr
\+ & S. Rajpoot, Mod. Phys. Letts. {\bf A7}, 309 (1992); \cr
\+ & D. S. Du and Z. Z. Xing, Phys. Rev. {\bf D 48}, 2349 (1993);\cr
\+ & E. Boridy, C. Hamzaoui, F. Lemay and
J. Lindig, unpublished report UQAM-PHE \cr
\+ & /94-10(Sept. 1994).\cr

\vglue2in

\noindent
{\bf Figure Caption:}
\smallskip
\noindent
Figure 1: A possible dominant mass generating mechanism which
causes factorization of Yukawa couplings related to mass
matrices.  The internal fermion and boson should be flavor
neutral.
\end